\documentclass[aps,pre,10pt,showpacs,floatfix,onecolumn,nofootinbib,a4paper]{revtex4-2}
\usepackage{amssymb, amsmath, amsthm, mathtools}
\usepackage{bm}
\usepackage{mathrsfs}
\usepackage{graphicx}
\usepackage{epsfig,color}
\usepackage{float}
\usepackage{subcaption}
\usepackage{verbatim}
\usepackage{dcolumn}
\usepackage{tikz}
\usepackage{url}

\usepackage[utf8x]{inputenc}
\usepackage[english]{babel}
\usepackage{mathrsfs}
\usepackage{epsfig}
\usepackage{listings}

\usepackage{mdwlist}
\usepackage{bm}
\usepackage{dsfont}
\usepackage{oldgerm}
\usepackage{geometry}
\usepackage{relsize}
\usepackage{afterpage}
\usepackage{lmodern}
\usepackage{tabularx}
\usepackage{cancel}

\newcommand{\dd}{\mathrm{d}}

\newcommand{\vt}{\vartheta}

\newcommand{\trans}{^\mathsf{T}}
\newcommand{\av}{\bm{a}}
\newcommand{\n}{\bm{n}}
\newcommand{\m}{\bm{m}}
\newcommand{\nper}{\bm{n}_\perp}
\newcommand{\mper}{\bm{m}_\perp}
\newcommand{\cv}{\bm{c}}

\newcommand{\ca}{\bm{c}^\ast}

\newcommand{\da}{\bm{d}^\ast}
\newcommand{\e}{\bm{e}}
\newcommand{\x}{\bm{x}}
\newcommand{\zero}{\bm{0}}
\newcommand{\y}{\bm{y}}
\newcommand{\nablatwo}{\nabla^2}
\newcommand{\normal}{\bm{\nu}}
\newcommand{\surface}{\mathscr{S}}

\newcommand{\nablas}{\nabla\!_\mathrm{s}}

\newcommand{\f}{\bm{f}}

\newcommand{\bv}{\bm{b}}

\newcommand{\F}{\mathbf{F}}
\newcommand{\I}{\mathbf{I}}
\newcommand{\Ln}{\mathbf{L}_{\n}}
\newcommand{\Lm}{\mathbf{L}_{\m}}

\newcommand{\C}{\mathbf{C}}
\newcommand{\B}{\mathbf{B}}

\newcommand{\curl}{\operatorname{curl}}
\newcommand{\tr}{\operatorname{tr}}

\newcommand{\Cf}{\C_{\f}}
\newcommand{\Cp}{\C_\phi}
\newcommand{\nay}{(\nabla\y)}
\newcommand{\nan}{(\nabla\normal)}
\newcommand{\curvature}{(\nablas\normal)}
\newcommand{\A}{\mathbf{A}}
\newcommand{\body}{\mathscr{B}}
\newcommand{\euclid}{\mathscr{E}}
\renewcommand{\frame}{(\e_1,\e_2,\e_3)}
\newcommand{\framem}{(\m,\mper,\e_3)}
\newcommand{\framen}{(\n,\nper,\normal)}
\newcommand{\ab}{c}
\newcommand{\slab}{\mathsf{S}}
\newcommand{\principal}{\bm{\tau}}
\newcommand{\curve}{\mathscr{C}}
\newcommand{\region}{\mathscr{A}}
\newcommand{\diver}{\operatorname{div}} 

\begin{document}
	\title{A blend of stretching and bending in nematic polymer networks}
	\author{Olivier Ozenda}
	\email{o.ozenda@unipv.it}
	\affiliation{Dipartimento di Matematica, Universit\`a di Pavia, Via Ferrata 5, 27100 Pavia, Italy}
	\author{Andr\'e M. Sonnet}
	\email{andre.sonnet@strath.ac.uk}
	\affiliation{Department of Mathematics and Statistics, University of Strathclyde, 26 Richmond Street, Glasgow G1 1XH, U.K. }
	\author{Epifanio G. Virga}
	\email{eg.virga@unipv.it}
	\affiliation{Dipartimento di Matematica, Universit\`a di Pavia, Via Ferrata 5, 27100 Pavia, Italy }

	\date{\today}

	\begin{abstract}
Nematic polymeric networks are 	(heat and light) activable materials, which combine the features of rubber and nematic liquid crystals.	
When only the stretching energy of a thin sheet of nematic polymeric network is minimized, the intrinsic (Guassian) curvature of the shape it takes upon (thermal or optical) actuation is determined. This, unfortunately, produces a multitude of possible shapes, for which we need a selection criterion, which may only be provided by a correcting bending energy depending on the extrinsic curvatures of the deformed shape. The literature has so far offered approximate corrections depending on the mean curvature. In this paper, we derive the appropriate bending energy of a sheet of polymeric nematic network from the celebrated neo-classical energy of nematic elastomers in three space dimensions. This task is performed via a dimension reduction based on a modified Kirchhoff-Love hypothesis, which withstands to the criticism of more sophisticated analytical tools. The result is a surface elastic free-energy density where stretching and bending are blended together; they may or may not be length-separated, and should be minimized together. The extrinsic curvatures of the deformed shape not only feature in the bending energy through the mean curvature, but also through the relative orientation of the nematic director in the frame of the directions of principal curvature.  
	\end{abstract}
	
	\maketitle

\section{Introduction}\label{sec:intro}
As wonderfully put in \cite{bhattacharya:material}, at small scales ``the material is the machine'', meaning that mechanical efficiency demands getting rid of shafts, pistons, gears, screws, and cogs and produce work by making the material itself change its shape. Thus, controllable changes of shape in thin and slender bodies actuated by external stimuli, which range from light to heat, and humidity, have become a new theme  of soft matter mechanics. A theme that Warner, in his recent review \cite{warner:topographic} which supplements White and Broer's \cite{white:programmable} (both highly recommended), calls \emph{topographic mechanics}.\footnote{Although---he says---``[we] might equally have called it \emph{metric mechanics}.''} Another possible name for it could be \emph{morphic mechanics}, which has the advantage of alluding directly to the biological\footnote{Huxley coined this adjective in 1955 as a derivative of \emph{morphism} \cite{huxley:morphism,huxley:heterosis} with a natural biological connotation, referring to the different forms in a species.} inspiration recently incarnated in a number of intriguing soft matter mechanics papers (see, for example, \cite{gladman:biomimetic,siefert:bio-inspired}).

Here we are concerned with liquid crystal solids, specifically with nematic polymeric networks. These materials are deformable like elastomers and highly responsive to stimuli like nematic liquid crystals. Their versatility arises from having rod-like molecules appended to the polymeric network constituting the elastomeric matrix. This feature imparts the anisotropy typical of nematic liquid crystals to the polymeric chain distribution; it is enough to open up a wealth of new scenarios for science and applications. Perhaps, the first striking manifestation of the opportunities offered by these materials was the swimmer ``that swims into the dark'' \cite{camacho-lopez:fast}, a  shell   flapping on a fluid surface activated by light.

We shall only consider thin shells of nematic polymeric networks. Both theory and experiments abound in this field,\footnote{The contributions most relevant to our development are duly referred to in the following sections.} but according to \cite{warner:topographic} a \emph{grand challenge} remains untouched, namely, the control of \emph{extrinsic bends}. As will also be recalled below, the Gaussian curvature of the deformed shell is entirely controlled by its metric, and so can reasonably be called \emph{intrinsic}. This, however, does not exhaust the curvature invariants; the mean curvature, for example, is missing, and being not dictated by the metric deserves the name of \emph{extrinsic}. A complete control of the deformed shell's shape can only result from the control of the extrinsic curvatures that feature in the bending elastic energy. As long as our scope is limited to the stretching elastic energy, as dominant as it may be, our vision will be blurred and our control incomplete. 

In this paper, we take up that challenge and we try and give a systematic answer to the problem of deriving the bending elastic energy density of a thin shell of nematic polymeric network from first principles.

For definiteness, our starting point in Sec.~\ref{sec:neo} will be the \emph{neo-classical} energy of nematic elastomers in bulk, but we deem our method sufficiently general to be applied to other more sophisticated expressions.

In Sec.~\ref{sec:stretching}, we review the stretching energy that has been so widely used in this field and see how its minimizers determine the Gaussian curvature of the deformed shape.

In Sec.~\ref{sec:connectors}, we introduce the notion of \emph{Cartesian connectors}, which in our view constitute a viable alternative to Christoffel symbols. In terms of these connectors, we reformulate the classical \emph{theorema egregium} of Gauss and the Codazzi-Mainardi compatibility conditions. 

Our method is a dimension reduction in the classical style of plate theory. Aware of the criticism moved against the Kirchhoff-Love hypothesis, which is a classical tool of that theory, in Sec.~\ref{sec:KLC}, we modify this hypothesis so as to make it compatible with recent, more sophisticated analytical tools, and put it at the basis of our method.

In Sec.~\ref{sec:blend}, which is the heart of the paper, we pull our strings together and derive the \emph{blended} energy formula, where stretching and bending conspire together, at different orders in the sheet's thickness.

Finally, in Sec.~\ref{sec:compatibility}, we consider the (usual) limit where stretching and bending energies are well-separated. We perform a \emph{two-step} minimization of the total elastic energy, regarding the bending energy as a perturbation to the leading stretching energy, and we use the former to select the optimal shape among the minimizers of the latter. It turns out that this (vanishing thickness) limit can only accommodate surfaces with positive Gaussian curvature (actually, only spheres), while surfaces with negative Gaussian curvature seem to be excluded.\footnote{We have reasons to conjecture this, but no full proof.} As an example, we show how a plane ribbon with a specific molecular orientation imprinted on it can be wrapped around a sphere by thermal (or optical) activation.

In Sec.~\ref{sec:conclusions}, we draw the conclusions of our work and pause briefly to consider the challenges that we still face.

The paper is closed by three appendices: in the first one we derive from the neo-classical energy the expression it takes when applied to nematic polymeric networks, in the second one   we substantiate our conjecture on the impossibility of surfaces with negative Gaussian curvature in the vanishing thickness limit, in the third one we show the numerical details employed in realizing the ribbon lifting.   

\section{Neo-classical energy}\label{sec:neo}
The theory of nematic elastomers was established in a series of works that started with \cite{blandon:deformation} and culminated in the book by Warner and Terentjev \cite{warner:liquid}, which still provides the most lucid and comprehensive introduction to the field. A vast literature has accompanied the birth of nematic elastomer science; we do not attempt to describe and comment on it, but some papers are particularly relevant to the theoretician and must be cited to establish a proper background for our study \cite{warner:soft,terentjev:orientation,verwey:soft,verwey:multistage,verwey:elastic,verwey:compositional}.

Perhaps, the landmark of all this endeavour is the formula for the stored energy for these anisotropic rubber materials, which is often, quite emphatically, named the ``trace formula''. In essence, this theory extends the principles governing the isotropic Gaussian distribution of polymeric chains to chains made anisotropic by the mutual interactions of nematogenic rod-like molecules appended to them. Not surprisingly then, the elastic free energy of the system (of a purely entropic nature) appears to be an extension of the classical neo-Hookian formula of isotropic rubber elasticity.\footnote{The neo-Hookian theory is a special case of the Mooney-Rivlin theory for rubber \cite[p.\,349]{truesdell:non-linear} (see also \cite{rivlin:large_I} for a mechanical derivation of the neo-Hookian formula). It has long been known that, despite (or maybe because of) its simplicity, the neo-Hookian theory is not generally adequate for ordinary rubber, except for very slow deformations \cite{rivlin:torsion}.} It is precisely this extended formula that represents the neo-classical energy; in this paper, it will be adopted as the starting point of our development, although the method that we shall illustrate can easily be applied to more elaborate expressions for the elastic free energy of nematic elastomers compatible with frame-indifference and symmetry requirements, such as those illuminated in \cite{desimone:elastic}.

It is not all bright. Since the very early days, criticism was not spared to the prolific literature on molecular theories for rubber elasticity \cite{truesdell:mechanical}, especially to the hypothesis about the Gaussian distribution of the long polymeric chains constituting the material.\footnote{More elaborate statistical theories were soon proposed in response to this criticism, see in particular \cite{wang:statistical} and \cite{treloar:non-Gaussian}. A comprehensive review of early molecular theories for rubber is the evergreen book \cite{treloar:non-linear_third}. Moreover, in recent years, models for the elastic response of entangled rubbers have been put forward, starting with Edward's \emph{tube} moldel \cite{edwards:theory}, after which a long list followed, including the \emph{localisation} model \cite{gaylord:localisation}, the \emph{slip-link} model \cite{ball:elasticity}, and the \emph{hoop} model \cite{higgs:trapped}.} Similar criticism could be moved to the trace formula for the elastic free energy of nematic elastomers, as it stems from assuming an anisotropic Gaussian distribution for polymeric chains. A noticeable improvement was achieved in \cite{kutter:tube} through a successful extension of Edward's tube model \cite{edwards:theory} for entangled rubber elasticity. Here, however, we shall abstain from dwelling on possible extensions of the trace formula, as we are merely interested  
in a method capable of producing from first principles the surface elastic free-energy density  of a thin sheet of nematic elastomer that incorporates both stretching and bending contributions; this method is not tied to any specific model for nematic rubber elasticity.

We now collect the ingredients that feature in the neo-classical formula. Nematic elastomers are isotropic solids with a fluid-like anisotropic ordering, which we describe classically by a \emph{director}. Two such director fields are needed to characterize the state of these materials, one for the \emph{reference} configuration, and one for the \emph{current} (deformed) configuration. We shall denote the former by $\m$ and the latter by $\n$. For definiteness, we may say that $\m$ describes the orientational order at the time of crosslinking and $\n$ that at the present time. More precisely, since the nematogenic molecular units are appended to the polymeric chains that constitute the elastic background  and are responsible for their anisotropic, albeit still Gaussian distribution in space, the appropriate measure of anisotropy is given by the polymer \emph{step length} tensor $\Lm$, which, following \cite{verwey:elastic} and \cite{nguyen:theory}, we write as 
 \begin{subequations}\label{eq:L_definitions}
 	\begin{equation}\label{eq:L_m}
 	\Lm:=a_0(\I+s_0\m\otimes\m)
 	\end{equation}
in the reference configuration, and 
\begin{equation}\label{eq:L_n}
\Ln:=a(\I+s\n\otimes\n)
\end{equation}
 \end{subequations}
in the current configuration. Here $\I$ is the identity (in  three-dimensional space), $a_0$ and $a$ are fixed positive parameters,\footnote{Corresponding to the persistence lengths perpendicular to the directions $\m$ and $\n$, respectively.} $s_0$ and $s$ are scalar measures of nematic order, defined as $s_0:=r_0-1$ and $s:=r-1$ in terms of the ratios of the principal chain step lengths (parallel/perpendicular to the corresponding directors in the reference and current configurations).

Letting $\body$ denote the reference configuration of a nematic elastomer  and $\f:\body\to\euclid$ a deformation of $\body$ in three-dimensional space $\euclid$, according to the neo-classical formula we write the soft elastic free energy density (per unit volume) of the body as 
\begin{equation}\label{eq:energy_density}
f_\mathrm{soft}:=\frac12\mu\tr(\F\trans\Ln^{-1}\F\Lm),
\end{equation}
where $\F:=\nabla\f$ is the deformation gradient, and $\mu>0$ is an elastic modulus, which scales linearly with temperature and is proportional to the number density of polymeric chains in the material. The energy density in \eqref{eq:energy_density} applies to ideal materials; non-ideal behaviours, such as  the semi-softness based on compositional fluctuations \cite{verwey:compositional} (see also \cite{biggins:semisoft} and \cite{nguyen:theory}) are ignored.

A major difference in the way nematic elastomers are produced is lucidly explained in \cite{white:programmable}: depending on the amount of crosslinking affecting the polymeric blend, the orientational and elastic degrees of freedom may be differently related. In the ideal, genuine nematic elastomer case, virtually no coupling exists between them, meaning that $\F$ and $\n$ can vary independently; this is the source of  the \emph{soft elastic modes} \cite{olmsted:rotational,verwey:elastic}, also envisaged in \cite{golubovic:nonlinear} on the basis of mere symmetry arguments. At the opposite end of the spectrum are \emph{nematic polymeric networks}, according to the name proposed in \cite{white:programmable};\footnote{Some also say that these are liquid crystal \emph{glasses} \cite{he:making,he:programmed,modes:disclination,plucinsky:programming}, while others prefer to say that they are nematic elastomers with a \emph{locked} (or \emph{frozen}) director \cite{cirak:computational}.} these are nematic elastomers in which the coupling between nematic order and network elasticity is complete. In nematic polymeric networks, the director $\m$ is \emph{blueprinted} in the elastic matrix \cite{modes:blueprinting} and $\m$ is conveyed into $\n$ by the deformation, that is, $\n$ is delivered by
\begin{equation}\label{eq:n}
\n=\frac{\F\m}{|\F\m|}.
\end{equation}   
In this paper, we shall only be concerned with nematic polymeric networks, and so \eqref{eq:n} will be enforced as a kinematic constraint. In general, elastomers are \emph{incompressible}, and so $\F$ will be subject to the further constraint
\begin{equation}\label{eq:incompressibility_F}
\det\F=1.
\end{equation}

With $\m$ imprinted in the reference configuration, we can still change the scalar order parameter, driving $s$ away from $s_0$ by either thermal or optical stimuli. For example, by heating the sample we can even destroy the nematic order present at the time of crosslinking, making $s=0$; this induces a contraction along $\m$ in the material (accompanied by a dilation in the orthogonal plane, to comply with the incompressibility constraint). Thus, work can be done and shape can be changed by light and heat. Needless to say, this opens fascinating scenarios for novel mechanical applications \cite{kowalski:pixelated,kowalski:voxel,babakhanova:liquid,zeng:light,brannum:deformation,van_oosten:glassy,van_oosten:bending,van_oosten:printed}. Here, for given $s_0>-1$, $s>-1$ will be the \emph{actuation parameter} of our theory. For definiteness, we shall assume that both $s_0$ and $s$ range in the interval $(-1,1)$, although the upper limit is not strictly necessary.

By use of \eqref{eq:L_definitions} and \eqref{eq:n}, we can express $f_\mathrm{soft}$ in \eqref{eq:energy_density} as
\begin{equation}\label{eq:F_definition}
f_\mathrm{soft}=\frac12\mu\frac{a_0}{a}F(\Cf),
\end{equation} 
where $\Cf:=\F\trans\F$ is the right Cauchy-Green tensor associated with the deformation $\f$, and
\begin{equation}\label{eq:bulk_energy_density}
F(\Cf)=\tr\Cf+\frac{s_0}{s+1}\m\cdot\Cf\m-\frac{s}{s+1}\frac{\m\cdot\Cf^2\m}{\m\cdot\Cf\m}.
\end{equation}
This formula, which delivers a frame-indifferent function of $\Cf$, has the same structure as equation (18) of \cite{cirak:computational}. (Its derivation is sketched in Appendix~\ref{sec:derivation}, for the reader's convenience.) It is the starting point of our development; in the following section, it will be subject to a naive dimension reduction with the aim of obtaining a first approximation to the elastic free-energy density (per unit area) for a thin sheet of nematic polymeric network.

\section{Stretching energy}\label{sec:stretching}
\begin{figure}[h]
\begin{center}
\includegraphics[width=.5\linewidth]{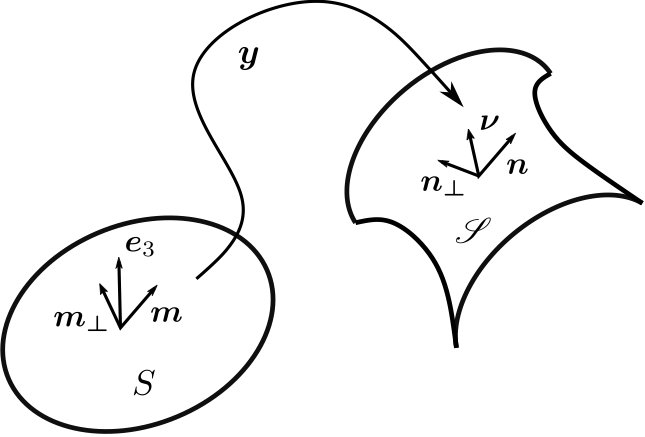}
\end{center}
\caption{\label{fig:sketch} The flat surface $S$ in the $(x_1,x_2)$ plane of a fixed Cartesian frame $\frame$ is deformed by the mapping $\y$ into a smooth surface $\surface$ embedded in three-dimensional Euclidean space $\euclid$. The blueprinted orientation is denoted by $\m$ in the reference configuration and by $\n$ in the current one; $\e_3$ is the outer unit normal to $S$, while $\normal$ is the outer unit normal to $\surface$;  correspondingly, $\mper:=\e_3\times\m$ and $\nper:=\normal\times\n$.}
\end{figure}

Consider a thin sheet of nematic polymeric network, which in its reference configuration occupies a flat region $S$ in the $(x_1,x_2)$ plane of a fixed Cartesian frame $(\e_1,\e_2,\e_3)$. Here we shall treat it as a two-dimensional inextensible membrane that can be deformed in  three-dimensional space $\euclid$, so as to take on the shape of a smooth surface $\surface$ (see Fig.~\ref{fig:sketch}).
Letting $\y:S\to\euclid$ denote a deformation of $S$ of class $C^2$, so that $\surface=\y(S)$, and taking $\m$ as a two-dimensional field on $S$, so that $\m\cdot\e_3\equiv0$, a general representation for $\nabla\y$ is the following,
\begin{equation}\label{eq:nabla_y_representation}
\nabla\y=\av\otimes\m+\bv\otimes\mper,
\end{equation}
where $\mper:=\e_3\times\m$. In \eqref{eq:nabla_y_representation},
both $\av$ and $\bv$ are vector fields defined on $S$, though they live in three-dimensional space $\euclid$ and are everywhere tangent to $\surface$. In the following, $\av$ and $\bv$ will be shorthands for $(\nabla\y)\m$ and $(\nabla\y)\mper$, respectively. It readily follows from \eqref{eq:nabla_y_representation} that
\begin{equation}\label{eq:C}
\C=(\nabla\y\trans)(\nabla\y)=a^2\m\otimes\m+\av\cdot\bv(\m\otimes\mper+\mper\otimes\m)+b^2\mper\otimes\mper,
\end{equation}
where $a^2:=\av\cdot\av$ and $b^2:=\bv\cdot\bv$.
The constraint of inextensibility for $S$  requires that $|\av\times\bv|=1$, and since 
\begin{equation}\label{eq:det_C}
\det\C=a^2b^2-(\av\cdot\bv)^2=|\av\times\bv|^2,
\end{equation}
we conclude that 
\begin{equation}\label{eq:det_C=1}
\det\C=1.
\end{equation}
Under this constraint, the unit normal to $\surface$ is
\begin{equation}\label{eq:normal}
\normal=\av\times\bv.
\end{equation}
In particular, it follows from \eqref{eq:nabla_y_representation} and \eqref{eq:normal} that
\begin{equation}\label{eq:nabla_y_property}
(\nabla\y)\trans\normal=\zero.
\end{equation}
Applying \eqref{eq:n} to the present setting, we may write $\av=a\n$ and define $\nper:=\normal\times\n$, so that the frame $\framen$ is oriented as the frame $\framem$. Since $\av$ and $\bv$ need not be orthogonal to one another, $\bv$ has in general components both along $\n$ and $\nper$.

The stretching tensor $\C$ in \eqref{eq:C} is a two-dimensional tensor defined on $S$; it obeys the identity
\begin{equation}\label{eq:Cayley_Hamilton}
\C^2=(\tr\C)\C-\I_2,
\end{equation}
where $\I_2$ is the identity in two space dimensions, as can be seen by combining the Cayley-Hamilton equation and \eqref{eq:det_C=1}.

A naive, but--as we shall see---effective dimension reduction would suggest to obtain the surface elastic free-energy density (scaled to $\frac12\mu\frac{a_0}{a}$) as
\begin{equation}\label{eq:pre_f_s}
f_s:=2hF(\C)=2h\frac{1}{s+1}\left(\tr\C+s_0\m\cdot\C\m+\frac{s}{\m\cdot\C\m}\right),
\end{equation}
where $F$ is given by \eqref{eq:bulk_energy_density} and $\C$ is the two-dimensional stretching tensor in \eqref{eq:C}. This is the energy associated with the \emph{stretching} of the membrane.

We now show that, for any given $\m$, $f_s$ in \eqref{eq:pre_f_s} attains its minimum on the constraint \eqref{eq:det_C=1} when the principal directions of $\C$ are $\m$ and $\mper:=\e_3\times\m$. To see this, we first compute
\begin{equation}\label{eq:dF_dC}
\frac{\partial F}{\partial\C}=\I+s_0\m\otimes\m-\frac{s}{(\m\cdot\C\m)^2}\m\otimes\m
\end{equation}
and
\begin{equation}\label{eq:d_detC_dC}
\frac{\partial\det\C}{\partial\C}=(\det\C)\C^{-1}.
\end{equation}
Then, requiring these derivatives to be proportional to one another, we easily conclude that for $\C$ to make $F$ stationary on \eqref{eq:det_C=1} it must be of the form
\begin{equation}\label{eq:C_stationary}
\C_0=\lambda_1^2\m\otimes\m+\lambda_2^2\mper\otimes\mper,
\end{equation}
where
\begin{equation}\label{eq:lambda_1_2}
\lambda_1:=\sqrt[4]{\frac{s+1}{s_0+1}}\quad\text{and}\quad\lambda_2=\frac1\lambda_1.
\end{equation}
Thus, for $s>s_0$, which is obtained upon cooling, the polymer network tends to extend along $\m$, whereas for $s<s_0$, which is obtained upon heating, the polymer network tends to extend along $\mper$.

Since
\begin{equation}\label{eq:d^2F_dC^2}
\frac{\partial^2F}{\partial\C^2}=\frac{2s}{(\m\cdot\C\m)^3}\m\otimes\m\otimes\m\otimes\m
\end{equation}
and $\m\cdot\C\m>0$, a standard convexity argument shows that $\C_0$ in \eqref{eq:C_stationary} is the unique minimizer of $F$ on the constraint \eqref{eq:det_C=1}.

A vast, beautiful literature is mostly concerned with finding the surfaces $\surface$ that, for a given field $\m$ imprinted on $S$, would comply with the required \emph{minimal stretches} in \eqref{eq:lambda_1_2}. A certainly incomplete list includes the papers \cite{modes:disclination,modes:gaussian,modes:negative,mostajeran:curvature,mostajeran:encoding,plucinsky:programming,mostajeran:frame,kowalski:curvature,warner:nematic}. Loosely, these studies have one feature in common: the attempt to characterize under different circumstances and with different methods the surfaces $\surface$ whose metric tensor is delivered by the optimal stretching tensor $\C_0$, as the meaning of the latter is precisely dictating how lengths of arcs on $S$ are supposed to be altered for the stretching energy $f_s$ to be minimized. Generally, this endeavour faces two distinct problems. First, there may or may not be a surface $\surface$ whose metric tensor is the optimal $\C_0$. Second, by Gauss' \emph{theorema egregium} (see, for example \cite[p.\,139]{stoker:differential}), a surface $\surface$ with a compatible, prescribed metric has a Gaussian curvature $K_0$ fully determined by $\C_0$ (and its spatial derivatives); this places a constraint on the admissible surfaces that may serve as stretching energy minimizers. Since the target Gaussian curvature $K_0$ depends on the imprinted director field $\m$, the latter problem turns eventually into the \emph{inverse} problem of devising $\m$ so as to realize a prescribed surface $\surface$, for a chosen value of the activation parameter $s$ \cite{griniasty:curved}.

As captivating as this programme may sound, the devil keeps lurking in the details, which here are multifaceted. Foremost all, is the multiplicity of shapes that can be realized by relaxing the (usual) $C^2$-smoothness requirement for the admissible surfaces $\surface$, allowing for example for sharp edges. Thus, a specific Gaussian curvature $K_0$ can be induced by solely minimizing the stretching energy \eqref{eq:pre_f_s}, but this is unlikely to select a single optimal shape for $\surface$, let aside the compatibility problem. To tackle this latter, means have been developed by what has recently come to be known as \emph{geometric elasticity} \cite{aharoni:geometry,aharoni:universal,griniasty:curved}. In essence, according to this theory, if the target metric associated with the stretching tensor $\C_0$ turns out to be geometrically incompatible, the optimal surface $\surface$ is determined by best approximating the target metric with the metric of an embeddable surface, in an appropriate $L^2$-norm. However, the issues of multiplicity of solutions and their smoothness remain unaffected.

Remedies to this state of affairs have already been proposed \cite{cirak:computational,mostajeran:encoding,plucinsky:actuation,krieger:tunable}; they mostly rely on the addition to $f_s$ of a correcting \emph{bending} energy, of one type or another, which penalizes extrinsic curvatures, thus providing a selection criterion that rules out the unwanted richness of minimizing shapes.\footnote{For thin sheets of isotropic polymeric networks subject to growth, similar models are also discussed in \cite{gemmer:shape_selection,gemmer:shape_transition}.} Despite their usefulness, these remedies suffer, however, from a certain degree of arbitrariness, as the proposed bending energies, as reasonable as they may be, are ultimately \emph{ad hoc}, and unrelated to the neo-classical formula for nematic elastomers \eqref{eq:energy_density}, or to any other model energy for rubber-like materials.\footnote{It should perhaps be noted that \cite{krieger:tunable} builds upon an early phenomenological model of de~Gennes  \cite{degennes:reflections} for nematic polymers (more recently also followed by \cite{uchida:elastic} and \cite{sawa:shape}), and so it is exempt from this critics. We shall see in Sec.~\ref{sec:blend} how we distance ourselves from this approach too.}  

Our approach in this paper is markedly different. By revisiting (and surpassing) the classical Kirchhoff-Love hypothesis,  much in the same spirit as Koiter's theory of thin shells \cite{koiter:consistent,koiter:nonlinear} (admirably revisited in \cite{steigmann:koiter}), we shall derive in Sec.~\ref{sec:blend} a theory for thin nematic polymeric networks in which stretching and bending energies may have the same order of magnitude and are thus treated on the same footing or, as we say, \emph{blended} together.  There, the stretching energy density in \eqref{eq:pre_f_s} will be fully justified.

In preparation for this, in the following section we refresh the preliminaries of differential geometry of surfaces in a way that avoids local charts of coordinates, but resorts instead to a number of vector fields, termed Cartesian \emph{connectors}, which describe the correspondence between local movable frames in the reference and current configurations of a material surface. 

\section{Cartesian connectors}\label{sec:connectors}
Here, for later use, we reformulate the essentials of the differential geometry of smooth surfaces embedded in three-dimensional space. For definiteness, we shall assume that the mapping $\y$ that deforms $S$ into $\surface$ is of class $C^3$. Likewise, the unit director field $\m$ imprinted on $S$ is assumed of class $C^2$, at least locally (along with its companion $\mper$). It should be kept in mind that both $\m$ and $\mper$ lie in the $(x_1,x_2)$ plane; $\nabla$ will denote the two-dimensional gradient in this plane, whereas $\nablas$ will denote the surface gradient on $\surface$.

The \emph{connector} $\cv$ is a vector field in the plane such that
\begin{subequations}\label{eq:connector_c}
	\begin{align}
	\nabla\m&=\mper\otimes\cv,\label{eq:connector_c_1}\\
	\nabla\mper&=-\m\otimes\cv.\label{eq:connector_c_2}
	\end{align}
\end{subequations}
The existence of $\cv$ and the specific form of \eqref{eq:connector_c} follow from the requirement that the pair of directors $(\m,\mper)$ be orthonormal everywhere on $S$.\footnote{The name \emph{connector} is inspired by the notion of \emph{spin connection} for surfaces (and manifolds) (see \cite{kamien:geometry}, for an effective introduction to the differential geometry useful in modelling soft matter).} Clearly, if $\m$ is known then $\cv$ is defined as $\cv:=(\nabla\m)\trans\mper$; on the other hand, if $\cv$ is assigned, at least locally, in the class $C^1$ then $\m$ (and $\mper$) can be determined up to a rigid rotation by solving equations \eqref{eq:connector_c}. To this end, however, $\cv$ must be \emph{compatible}; it follows from the symmetry of both $\nablatwo\m$ and $\nablatwo\mper$ that the compatibility condition reads as
\begin{equation}\label{eq:symmetry_grad_c}
\nabla\cv-(\nabla\cv)\trans=\zero,
\end{equation}
which for a simply connected $S$ implies that $\cv=\nabla\Phi$, where $\Phi$ is an appropriate scalar potential. Since we assume that both $\m$ and $\mper$ are known from the start, we shall here consider  $\cv$ as known and satisfying \eqref{eq:symmetry_grad_c}.

Combining the general representation for $\nabla\y$ in \eqref{eq:nabla_y_representation} and the requirement, akin to \eqref{eq:n}, that
\begin{equation}\label{eq:n_conveyed}
\n=\frac{(\nabla\y)\m}{|(\nabla\y)\m|},
\end{equation}
we can write $\av=a\n$ and $\bv=b'\n+b''\nper$. With these representations, $\av\cdot\bv=ab'$ and $\av\times\bv=ab''\normal$, and so with \eqref{eq:normal} we obtain 
\begin{equation}
b'=\frac1a\av\cdot\bv,\qquad b''=\frac1a.
\end{equation}

In complete analogy to the frame $\framem$ on $S$, 
we  describe the corresponding frame $\framen$ as a field of orthonormal directors on $\surface$. Equations \eqref{eq:connector_c} are generalized to
\begin{subequations}\label{eq:connectors_c_d}
	\begin{align}
	\nabla\n&=\nper\otimes\ca+\normal\otimes\da_1,\label{eq:nabla_n}\\
	\nabla\nper&=-\n\otimes\ca+\normal\otimes\da_2,\label{eq:nabla_n_perp}\\
	\nabla\normal&=-\n\otimes\da_1-\nper\otimes\da_2\label{eq:nabla_nu},
	\end{align}
\end{subequations}
where the connectors $\ca$, $\da_1$, and $\da_2$ are planar fields defined on $S$. A number of consequences for these fields follow from the integrability  condition that requires the second gradients of $\y$, $\n$, $\nper$, and $\normal$ to be symmetric: they are listed below.
\begin{enumerate}
	\item For the symmetry of $\nablatwo\y$,
	\begin{subequations}
		\begin{align}
	a\cv\cdot\m-\nabla a\cdot\mper+b'\cv\cdot\mper+\nabla b'\cdot\m-\frac1a\ca\cdot\m&=0	,\label{eq:c_star_m_perp}\\
	a\ca\cdot\mper-b'\ca\cdot\m-\frac1a\cv\cdot\mper+\frac{1}{a^2}\nabla a\cdot\m&=0,\label{eq:c_star_m}\\
a\da_1\cdot\mper-b'\da_1\cdot\m-\frac1a\da_2\cdot\m	&=0.\label{eq:d_12_d21}
		\end{align}
	\end{subequations}
In particular, \eqref{eq:c_star_m_perp} and \eqref{eq:c_star_m} can be combined together to yield 
\begin{equation}\label{eq:c_star}
\ca=\C\cv-\frac12\frac{\ab}{a^2}\nabla a^2+(\nabla\ab\cdot\m)\m+\frac12(\nabla b^2\cdot\m)\mper-\frac12(\nabla a^2\cdot\mper)\m,
\end{equation}
where we have set $\ab:=\av\cdot\bv$. By recalling \eqref{eq:C}, it becomes apparent from \eqref{eq:c_star} that $\ca$ is completely determined by $\cv$ and $\C$.\footnote{This latter is also contributing through the derivatives of its components in the basis $(\m,\mper)$, which are $C_{11}=a^2$, $C_{22}=b^2$, and $C_{12}=C_{21}=\ab$.}
	\item For the symmetry of $\nablatwo\n$,
	\begin{subequations}
		\begin{align}
		\nabla\ca-(\nabla\ca)\trans&=\da_1\otimes\da_2-\da_2\otimes\da_1,\label{eq:skew_part_nabla_c_ast}\\
		\nabla\da_1-(\nabla\da_1)\trans&=\da_2\otimes\ca-\ca\otimes\da_2,\label{eq:skew_part_nabla_d_1}
		\end{align}
	\end{subequations}
	which can also be written in the equivalent forms
	\begin{subequations}
		\begin{align}
		\curl\ca&=\da_2\times\da_1,\label{eq:skew_part_nabla_c_ast_equiv}\\
		\curl\da_1&=\ca\times\da_2.\label{eq:skew_part_nabla_d_1_equiv}
		\end{align}
	\end{subequations}
	\item For the symmetry of $\nablatwo\nper$, \eqref{eq:skew_part_nabla_c_ast} is supplemented by
	\begin{equation}\label{eq:skew_part_d_2}
	\nabla\da_2-(\nabla\da_2)\trans=\ca\otimes\da_1-\da_1\otimes\ca,
	\end{equation}
	or its equivalent form
	\begin{equation}\label{eq:skew_part_d_2_equiv}
	\curl\da_2=\da_1\times\ca.
	\end{equation}
	\item Finally, the symmetry of $\nablatwo\normal$ is guaranteed by \eqref{eq:skew_part_nabla_d_1} and \eqref{eq:skew_part_d_2}.
\end{enumerate} 
The connectors $\da_1$ and $\da_2$ can be given a geometric interpretation by computing the curvature tensor $\nablas\normal$ of $\surface$.
It readily follows from \eqref{eq:nabla_y_representation} that 
\begin{equation}\label{eq:nabla_y_inverse}
(\nabla\y)^{-1}=\frac1a\m\otimes\n+a\mper\otimes\nper-b'\m\otimes\nper.
\end{equation}
Letting
\begin{equation}\label{eq:d_representation}
\da_1=d_{11}\m+d_{12}\mper,\qquad\da_2=d_{21}\m+d_{22}\mper,
\end{equation}
from \eqref{eq:nabla_y_inverse} we arrive at
\begin{align}\label{eq:curvature_tensor}
\nablas\normal&=(\nabla\normal)(\nabla\y)^{-1}\nonumber\\
&=-\left[\frac{d_{11}}{a}\n\otimes\n+(ad_{12}-b'd_{11})\n\otimes\nper+\frac{d_{21}}{a}\nper\otimes\n+(ad_{22}-b'd_{21})\nper\otimes\nper\right],
\end{align}
which is duly symmetric, as by \eqref{eq:d_representation} equation \eqref{eq:d_12_d21} reduces to
\begin{equation}\label{eq:curvature_symmetry}
\frac1ad_{21}=ad_{12}-b'd_{11}.
\end{equation}
Both the mean curvature $H$ and the Gaussian curvature $K$ of $\surface$ can easily be derived from \eqref{eq:curvature_tensor}; they are given by
\begin{align}
H&=-\frac12\left(\frac{d_{11}}{a}+ad_{22}-b'd_{21}\right),\label{eq:H}\\
K&=d_{11}d_{22}-d_{12}d_{21}.\label{eq:K}
\end{align}

An important conclusion follows by combining \eqref{eq:skew_part_nabla_c_ast} and \eqref{eq:K} with the aid of \eqref{eq:d_representation}, namely
\begin{equation}\label{eq:nabla_c_star_skew_part}
\nabla\ca-(\nabla\ca)\trans=K(\m\otimes\mper-\mper\otimes\m).
\end{equation}
Since, as shown by \eqref{eq:c_star}, the left-hand side of \eqref{eq:nabla_c_star_skew_part} is determined by $\C$ (alongside the first and second derivatives of its components) and $\m$, so is $K$. In other words, the metric on $\surface$ and the imprinted director $\m$ determine the Gaussian curvature of $\surface$. This is the manifestation in our setting of the celebrated \emph{theorema egregium} of Gauss.\footnote{Similarly, equations \eqref{eq:skew_part_nabla_d_1} and \eqref{eq:skew_part_d_2} are related to the Codazzi-Mainardi equations (see, for example, \cite[p.\,144]{stoker:differential}).}

In the special case where $\C$ is $\C_0$, by comparing \eqref{eq:C_stationary} and \eqref{eq:C}, we see that $a^2=\lambda_1^2$, $b^2=\frac{1}{\lambda_1^2}$, and so $b'=0$ and $\ab=0$, with $\lambda_1$ the (constant) principal stretch given by \eqref{eq:lambda_1_2}. In this case, \eqref{eq:curvature_tensor} and \eqref{eq:c_star} reduce to
\begin{equation}\label{eq:curvature_tensor_reduced}
\nablas\normal=-\left(\frac{d_{11}}{\lambda_1}\n\otimes\n+\lambda_1d_{12}\n\otimes\nper+\frac{d_{21}}{\lambda_1}\nper\otimes\n+\lambda_1d_{22}\nper\otimes\nper\right)
\end{equation}
and
\begin{equation}\label{eq:c_star_reduced}
\ca=\lambda_1^2(\cv\cdot\m)\m+\frac{1}{\lambda_1^2}(\cv\cdot\mper)\mper,
\end{equation}
respectively, while 
\eqref{eq:curvature_symmetry} becomes
\begin{equation}\label{eq:d_12_identity}
\lambda_1d_{12}=\frac{d_{21}}{\lambda_1}.
\end{equation} It follows 
from \eqref{eq:c_star_reduced} that
\begin{equation}\label{eq:nabla_c_star_skew_part_pre}
\nabla\ca-(\nabla\ca)\trans=\left(\lambda_1^2-\frac{1}{\lambda_1^2}\right)(c_2^2-c_1^2+c_{12})(\m\otimes\mper-\mper\otimes\m),
\end{equation}
where we have set
\begin{equation}\label{eq:c1_c2_c12}
c_1=\cv\cdot\m,\quad c_2=\cv\cdot\mper,\quad c_{12}=\m\cdot(\nabla\cv)\mper.
\end{equation}
A comparison with \eqref{eq:nabla_c_star_skew_part} readily leads us to conclude that 
\begin{equation}\label{eq:theorema_egregium_reduced}
K=\left(\lambda_1^2-\frac{1}{\lambda_1^2}\right)(c_2^2-c_1^2+c_{12}).
\end{equation}
It is not difficult to check that \eqref{eq:theorema_egregium_reduced} is precisely equation (22) of \cite{mostajeran:curvature}, which was deduced with the more traditional use of coordinates and Christoffel symbols. We shall need \eqref{eq:theorema_egregium_reduced} when in Sec.~\ref{sec:compatibility} we shall consider the vanishing thickness limit of our blended energy.

\section{Modified Kirchhoff-Love hypothesis}\label{sec:KLC}
\begin{figure}[h]
	\begin{center}
		\includegraphics[width=.6\linewidth]{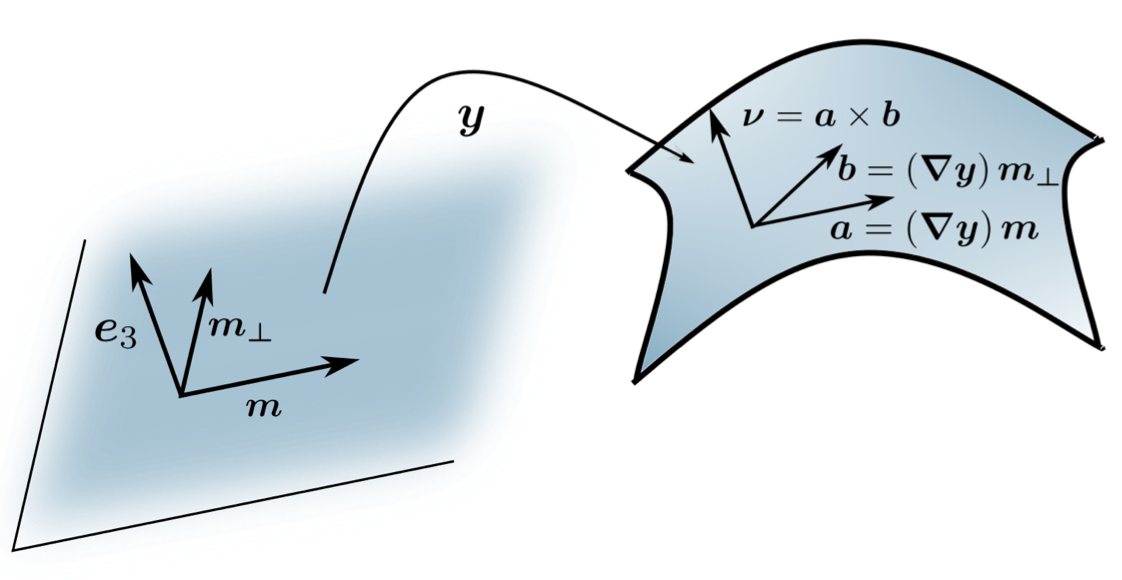}
	\end{center}
	\caption{\label{fig:scheme}
	A thin sheet of nematic polymeric network of thickness $2h$ on which the director $\m$, imprinted as in Fig.~\ref{fig:sketch} on the mid surface $S$ in the reference configuration, is extended uniformly across the cross-section. The mapping $\y$ deforms $S$ into the smooth surface $\surface$ as illustrated in Sec.~\ref{sec:stretching}; the director field $\n$, conveyed on $\surface$ by $\nabla\y$ as in \eqref{eq:n_conveyed}, is extended uniformly along the normal $\normal$ across the cross-section in the current configuration.}
\end{figure}

Here and in the following, we shall represent a thin sheet of nematic polymeric network in its reference configuration as the region in  three-dimensional space defined by $\slab:=\{(\x,x_3): \x\in S, -h\leqq x_3\leqq h\}$; it is a flat slab of thickness $2h$ around the mid surface $S$, as shown in Fig.~\ref{fig:scheme}.
The nematic director $\m$ is taken to be imprinted so as to be uniform across the sheet's cross-section; it suffices to know its trace on $S$ to know it throughout the whole sheet. Likewise, the deformed mid surface $\surface$ is obtained by the  mapping $\y$, as in Sec.~\ref{sec:stretching}, the nematic director $\m$ is conveyed into $\n$ on $\surface$ according to \eqref{eq:n_conveyed} and then extended uniformly along the outer normal $\normal$ across the cross-section of the current configuration of the sheet. The general representation for $\nabla\y$ in \eqref{eq:nabla_y_representation} still applies and $\normal$ is given by \eqref{eq:normal}.

The Kirchhoff-Love hypothesis in classical shell theory assumes that any point situated on a normal to the mid surface $S$ remains, after the deformation has taken place, on the corresponding normal to the deformed surface $\surface$, and at the \emph{same distance} from it (see, for example, \cite[p.\,551]{villaggio:mathematical} or \cite[p.\,156]{ciarlet:introduction}). As a consequence, the surfaces parallel to $S$ that bound the sheet in the reference configuration (at the distance $2h$ from one  another) are deformed into surfaces parallel to $\surface$ in the current configuration (with unchanged mutual distance); the same could be said for any intermediate surface parallel to $S$ in the reference configuration. This hypothesis has a long and controversial history, which also casts a shadow on its correct attribution.\footnote{We can only afford here a few passing remarks to substantiate this statement. One treatise \cite[p.\,5]{libai:nonlinear} attributes it to Kirchhoff and puts it at the basis of their theory of shells with the dignity of a constitutive assumption (instead of a humbler kinematic hypothesis), another \cite[p.\,551]{villaggio:mathematical} questions that Love had any role in advancing this hypothesis, as his original shell theory, as recounted in \cite[Art.\,329]{love:treatise}, allegedly ``rests on other, less convincing, assumptions''. It has also been argued \cite{friesecke:theorem} that this hypothesis has wrongly been associated with Kirchhoff's theory of shells. It should rather be more appropriate to treat it as an ansatz typical of the Cosserat intrinsic theory of shells \cite{cosserat:theorie,cosserat:theorie_livre}, accurately presented in the book \cite[Sec.~XIV.13]{antman:nonlinear}.}

We shall \emph{not} use this hypothesis here; rather, we modify it. One reason why the traditional Kirchhoff-Love hypothesis (we still call it so) may be considered inappropriate was illustrated in \cite{friesecke:theorem}: it was found by a rigorous application of the $\Gamma$-convergence method that   the limiting  bending energy (for vanishing thickness) of an isotropic elastic plate does \emph{not} fully agree with that delivered by the kinematic restriction envisaged by the Kirchhoff-Love hypothesis. The $\Gamma$-limit energy would instead be compatible with a different pattern, which includes stretches along the plate's mid surface. It is shown in \cite{ozenda:kirchhoff} that the modified Kirchhoff-Love hypothesis proposed here reproduces exactly the $\Gamma$-limit obtained in \cite{friesecke:theorem}. Reassured by this comparison, we now describe our method.

We assume that all points of $\slab$ suffer a deformation described by
\begin{equation}\label{eq:deformation}
\f(\x+x_3\e_3)=\y(\x)+\phi(\x,x_3)\normal(\x),
\end{equation}
where $\x$ ranges in $S$, $-h\leqq x_3\leqq h$,  $\normal$ is the unit outer normal to $\surface$, expressed as a function of $\x$, and $\phi$ is a smooth scalar function to be determined so as to ensure that $\f$ obeys the incompressibility constraint (in three space dimensions), granted that $S$ is inextensible (in two space dimensions).

In accord with the notation used so far, we shall denote by $\nabla\phi$ the gradient of $\phi$ in $\x$ and by $\phi'$ the partial derivative of $\phi$ with respect to $x_3$. Clearly, $\phi$ must satisfy the following condition, 
\begin{equation}\label{eq:phi_0_condition}
\phi(\x,0)\equiv0,
\end{equation}
which is a consequence of \eqref{eq:deformation}. As shown in \cite{ozenda:kirchhoff}, the local invertibility of $\f$ demands that $\phi'(\x,0)\neq0$, and so we can assume that $\phi'(\x,0)>0$ for all $\x\in S$. Since \eqref{eq:phi_0_condition} implies that $\nabla\phi(\x,0)\equiv\bm{0}$, there is a sufficiently small $h$ such that 
\begin{equation}\label{eq:kinematic_assumption}
|\nabla\phi|\ll\phi'.
\end{equation}
In the following, it will be understood that $h$ is such that \eqref{eq:kinematic_assumption} applies.
Moreover, requiring $\phi$ to be of class $C^3$, we shall also use for it the following polynomial representation,
\begin{equation}\label{eq:phi_polynomial}
\phi(\x,x_3)=\alpha(\x)x_3+\beta(\x)x_3^2+\gamma(\x)x_3^3+O(x_3^4),
\end{equation}
which is consistent with \eqref{eq:kinematic_assumption} in view of the regularity required for the functions $\alpha$, $\beta$, and $\gamma$.

It follows from \eqref{eq:deformation} that the deformation gradient reads as
\begin{equation}\label{eq:defomation_gradient}
\F:=\mathrm{D}\f=\nabla\y+\phi\nabla\normal+\phi'\normal\otimes\e_3+\normal\otimes\nabla\phi.
\end{equation}
A direct computation using \eqref{eq:nabla_y_representation} and    \eqref{eq:kinematic_assumption} shows that the right Cauchy-Green tensor $\Cf$ associated with $\f$ can be written in the following form,
\begin{equation}\label{eq_C_f}
\Cf:=\F\trans\F=\Cp+\phi'^2\e_3\otimes\e_3 +o(\phi'^2)
\end{equation}
where 
\begin{equation}\label{eq:C_phi}
\Cp:=\C+\phi\C_1+\phi^2\C_2
\end{equation}
with
\begin{equation}\label{eq:C_1_C_2}
\C_1:=\nay\trans\nan+\nan\trans\nay\quad\text{and}\quad\C_2:=\nan\trans\nan.
\end{equation}
$\Cp$, like $\C$, is a two-dimensional tensor, and so it satisfies
\begin{equation}\label{eq:Cayley}
\Cp^2-(\tr\Cp)\Cp+(\det\Cp)\I_2=\zero.
\end{equation}
Moreover, since
\begin{equation}\label{eq:curvature_tensor_old}
\nabla\normal=\curvature\nay,
\end{equation}
where $\nablas\normal$ is the curvature tensor of $\surface$, $\C_1$ and $\C_2$ can be given the following equivalent forms,
\begin{equation}\label{eq:C_1_C_2_alternative}
\C_1=2\nay\trans\curvature\nay\quad\text{and}\quad\C_2=\nay\trans\curvature^2\nay.
\end{equation}

The constraint of bulk incompressibility requires that
\begin{equation}\label{eq:incompressibility}
\det\Cf=\phi'^2\det\Cp=1.
\end{equation}
This  equation subject to \eqref{eq:phi_0_condition} determines $\phi$. To solve it, we first compute $\det\Cp$ up to second order in $\phi$, and then we make use of \eqref{eq:phi_polynomial}. By use of the identities 
\begin{equation}\label{eq:A_identities}
\det\A=\A\e_1\times\A\e_2\cdot\e_3\quad\text{and}\quad\tr\A=(\A\e_1\times\e_2+\e_1\times\A\e_2)\cdot\e_3,
\end{equation}
valid for all two-dimensional  tensors $\A$,
one obtains that
\begin{equation}\label{eq:det_C_phi}
\det\Cp=1+\phi\tr(\C^{-1}\C_1)+\phi^2\tr(\C^{-1}\C_2)+\phi^2\det\C_1+O(\phi^3),
\end{equation}
which in view of \eqref{eq:C_1_C_2_alternative} can be given the form
\begin{equation}\label{eq:det_C_phi_curvature}
\det\Cp=1+4H\phi+2(2H^2+K)\phi^2+O(\phi^3).
\end{equation}
Equating equal powers of $x_3$ in equation  \eqref{eq:incompressibility}, with the aid of \eqref{eq:phi_polynomial}, we determine $\alpha$, $\beta$, and $\gamma$, thus arriving at
\begin{equation}\label{eq:phi_powers}
\phi=x_3-Hx_3^2+\frac13(6H^2-K)x_3^3+O(x_3^4).
\end{equation}
This equation shows clearly how we depart here from the classical Kirchhoff-Love hypothesis; had we adopted it, $\phi$ would be $\phi\equiv x_3$.\footnote{This is the case for equation (3) of \cite{krieger:tunable}, where, in a spirit similar to ours, a dimensional reduction is performed for nematic elastomers treated within de~Gennes' phenomenological model.}

In particular, it follows from \eqref{eq:phi_powers} that the thickness $2h'$ of the deformed sheet is given by
\begin{equation}\label{eq:thickness}
2h'=\int_{-h}^{+h}\phi'\dd x_3=2h +\frac23h^3(6H^2-K)+O(h^5),
\end{equation}
and so it is not uniform across the deformed sheet.
Three more formulas similar to \eqref{eq:thickness} follow from \eqref{eq:phi_powers}; we record them here for completeness, 
\begin{equation}\label{eq:phi_integrals}
\int_{-h}^{+h}\phi'^2\dd x_3=2h+\frac43h^3(8H^2-K),\quad\int_{-h}^{+h}\phi\dd x_3=-\frac23h^3H,\quad\int_{-h}^{+h}\phi^2\dd x_3=\frac23h^3,
\end{equation}
all valid to within terms $O(h^5)$. 

In the following section, we shall apply the representation in  \eqref{eq:deformation} for the deformation $\f$ with $\phi$ as in \eqref{eq:phi_powers}, to develop a dimension reduction strategy that delivers up to terms in $h^3$ the elastic free energy stored in thin nematic polymeric networks according to the neo-classical formula recalled in Sec.~\ref{sec:neo}.

\section{Energy blending}\label{sec:blend}
Our strategy will simply be to apply the method outlined in the preceding section to the (scaled) elastic free-energy density (per unit volume) in \eqref{eq:bulk_energy_density}.

By use of \eqref{eq_C_f}, \eqref{eq:Cayley}, and \eqref{eq:incompressibility} in \eqref{eq:bulk_energy_density}, we first give $F(\Cf)$ a more compact form,
\begin{equation}\label{eq:energy_density_Cp}
F(\Cf)=\frac{1}{\det\Cp}+\frac{1}{s+1}\left\{\tr\Cp+s_0\m\cdot\Cp\m+s\frac{\det\Cp}{\m\cdot\Cp\m}\right\}.
\end{equation}
The desired elastic free-energy density $f$ per unit area of $S$ will finally be obtained by expanding  $F(\Cf)$ up to second order in $\phi$ and then integrating in $x_3$ over the thickness of the sheet. The result will be written in the form
\begin{equation}\label{eq:energy_splitting}
f(\nabla\y,\nabla^2\y,\m):=\int_{-h}^hF(\Cf)\dd  x_3=hf_1+h^3f_3+O(h^5),
\end{equation}
where $hf_1$ is $O(h)$ and represents the \emph{stretching} energy, whereas $h^3f_3$ is $O(h^3)$ and represents the \emph{bending} energy.

To perform this task, two identities turned out to be useful. We now  proceed to prove the first. Let
\begin{equation}\label{eq:a_1_a_2}
a_1:=\av\cdot\curvature\av\quad\text{and}\quad a_2:=\av\cdot\curvature^2\av,
\end{equation}
so that , by \eqref{eq:C_1_C_2_alternative}, we have that $\m\cdot\C_1\m=2a_1$ and $\m\cdot\C_2\m=a_2$. Since $\nablas\normal$ can be represented as
\begin{equation}\label{eq:curvature_tensor_representation}
\nablas\normal=\kappa_1\principal_1\otimes\principal_1+\kappa_2\principal_2\otimes\principal_2,
\end{equation}
where $\kappa_1$, $\kappa_2$ are the principal curvatures of $\surface$ and $\principal_1$, $\principal_2$ the corresponding principal directions of curvature (oriented so that $\normal=\principal_1\times\principal_2$), we readily obtain that 
\begin{equation}\label{eq:a_1_a_2_new}
a_1=\kappa_1(\av\cdot\principal_1)^2+\kappa_2(\av\cdot\principal_2)^2\quad\text{and}\quad a_2=\kappa_1^2(\av\cdot\principal_1)^2+\kappa_2^2(\av\cdot\principal_2)^2,
\end{equation}
from which it follows that 
\begin{equation}\label{eq:a_identity}
2Ha_1=a_2+Ka^2,
\end{equation}
showing how $a_1$ and $a_2$ are related. By \eqref{eq:C_phi}, we then conclude that
\begin{equation}\label{eq:C_phi_m_m}
\m\cdot\Cp\m=a^2(1-K\phi^2)+2\phi a_1(1+\phi H),
\end{equation}
which is needed in \eqref{eq:energy_density_Cp}.

In a similar way, we compute $\tr\Cp$, making use of \eqref{eq:C_phi} and \eqref{eq:C_1_C_2_alternative},
\begin{equation}\label{eq:tr_C_phi}
\tr\Cp=\tr\C+2\phi b_1+\phi^2b_2,
\end{equation}
where we have set
\begin{equation}\label{eq:b_1_b_2}
b_1:=\tr(\B\nablas\normal)\quad\text{and}\quad b_2:=\tr\left(\B\curvature^2\right),	
\end{equation}
and $\B:=(\nabla\y)(\nabla\y)\trans$ is the left Cauchy-Green tensor associated with the deformation $\y$. The second useful identity is the following,
\begin{equation}
\label{eq:second_identity}
2Hb_1=b_2+K\tr\C,
\end{equation}
which is an immediate consequence of \eqref{eq:curvature_tensor_representation} and the fact that $\tr\B=\tr\C$. Inserting \eqref{eq:second_identity} in \eqref{eq:tr_C_phi}, we obtain
\begin{equation}\label{eq:tr_C_phi_computed}
\tr\Cp=\tr\C(1-K\phi^2)+2\phi b_1(1+\phi H),
\end{equation}
which parallels \eqref{eq:C_phi_m_m}.

It is now a matter of (lengthy, but easy) computations to establish that
\begin{equation}\label{eq:f_s}
f_1=2\Big\{1+\frac{1}{s+1}\Big[\tr\C+s_0\m\cdot\C\m+\frac{s}{\m\cdot\C\m}\Big]\Big\},	
\end{equation}
\begin{equation}\label{eq:f_b}
f_3=\frac23\left\{2(8H^2-K)+\frac{1}{s+1}\left[\left(\frac{3s}{a^2}-a^2s_0-\tr\C\right)K-\frac{4s}{a^2}(2H-\hat{a}_1)\hat{a}_1\right] \right\},
\end{equation}
where\footnote{The reader should not be surprised by not seeing $b_1$ in either \eqref{eq:f_s} or \eqref{eq:f_b}; it only features in the coefficient of the $x_3$-linear term  in the integrand of \eqref{eq:energy_splitting}, and so eventually it does not contribute to the integral.}
\begin{equation}\label{eq:a_1_hat}
\hat{a}_1:=\frac{a_1}{a^2}=\n\cdot\curvature\n
\end{equation}
and we recall that $a^2=\m\cdot\C\m$.
A number of remarks are in order.

First, comparing \eqref{eq:f_s} and \eqref{eq:pre_f_s}, we readily see that 
\begin{equation}\label{eq:f_1}
hf_1=2h+f_s,
\end{equation}
which fully justifies taking $f_s$ as stretching energy-density, as it differs from $hf_1$ only by an inessential additive constant. Thus, setting $f_b:=h^3f_3$ for the \emph{bending} energy, we shall hereafter write $f$ in \eqref{eq:energy_splitting} as $f=f_s+f_b+O(h^5)$.

Second, as is apparent from \eqref{eq:f_s}, the stretching energy $f_s$ depends only on $\C$ and $\m$, whereas the bending energy $f_b$ also depends on invariant measures of extrinsic curvature of $\surface$ and specifically, via $\hat{a}_1$, on the orientation of the conveyed director $\n$ in the frame of the principal directions of curvature.

Third, by representing $\n$ as
\begin{equation}\label{eq:n_representation}
\n=\cos\varphi\principal_1+\sin\varphi\principal_2,
\end{equation}
where $-\frac\pi2\leqq\varphi\leqq\frac\pi2$ is the angle that the director makes with the principal direction of curvature $\principal_1$,
standard computations deliver the following formula
\begin{equation}\label{eq:energy_blended}
\begin{split}
\frac{1}{2h}(f_s+f_b)=1&+\frac{1}{s+1}\left(b^2+a^2+s_0a^2+\frac{s}{a^2}\right)\\
+&\frac13h^2\bigg\{K\left[6-\frac{1}{s+1}\left(\frac{s}{a^2}+s_0a^2+a^2+b^2\right)+\frac{2s}{(s+1)a^2}\sin^22\varphi\right]
\\ +&2C\left(4-\frac{s}{(s+1)a^2}\sin^22\varphi\right)\bigg\},
\end{split}
\end{equation}
where $b^2=\mper\cdot\C\mper$ and $C:=\frac12(\kappa_1^2+\kappa_2^2)$, which is also referred to as the Casorati curvature of $\surface$ \cite{casorati:mesure}. This formula will be expedient in the following section, where we inquire how to minimize $f$ when $f_b$ can be considered as a perturbation to $f_s$, that is, when stretching and bending take place on separate length scales.

\section{Vanishing thickness limit}\label{sec:compatibility}
We study in this section the limit in which $f_s$ and $f_b$ are scale-separated, with $f_s$ prevailing over $f_b$. This means that we can envision a sort of \emph{two-step} minimization strategy according to which we first minimize $f_s$ and, on its minimizers, we further minimize $f_b$. If this strategy has the advantage of reducing the abundance of shapes with minimal stretching energy already lamented in Sec.~\ref{sec:stretching}, it risks clashing with the geometric compatibility conditions in the large, which in the language of connectors are embodied by  \eqref{eq:skew_part_nabla_c_ast} (\emph{theorema egregium}) and \eqref{eq:skew_part_nabla_d_1}, \eqref{eq:skew_part_d_2} (Codazzi-Mainardi conditions). Here, we resolve this geometric issue (almost completely). In particular, it will be clear how this approximate strategy differs from the classical approach of minimizing over isometries only.

First, we know from Sec.~\ref{sec:stretching} that minimizing $f_s$ amounts to set $\C=\C_0$ as in \eqref{eq:C_stationary}, which prescribes the Gaussian curvature $K$ as in \eqref{eq:theorema_egregium_reduced}, which we rewrite as
\begin{equation}\label{eq:constraint_K}
K=\left(\lambda_1^2-\frac{1}{\lambda_1^2}\right)\kappa,	 	 
\end{equation} 
having set
\begin{equation}
\label{eq:kappa}
\kappa:=c_2^2-c_1^2+c_{12},
\end{equation}
see also \eqref{eq:c1_c2_c12}. We are thus back to isometries, but with further ammunition: the minimization over what still remains free in the bending energy. With this in mind, we express $f_b$ in the form
\begin{equation}\label{eq:f_b_iper_reduced}
f_b=\frac23h^3\left\{2K\left(3-\frac{1}{\lambda_1^2}+\frac{s}{s+1}\frac{1}{\lambda_1^2}\sin^22\varphi \right)+2C\left(4-\frac{s}{s+1}\frac{1}{\lambda_1^2}\sin^22\varphi\right)\right\},
\end{equation}
where we have set $a^2=$ and $b^2=\frac{1}{\lambda_1^2}$ and $\lambda_1$ is given by \eqref{eq:lambda_1_2}. In \eqref{eq:f_b_iper_reduced}, $K=\kappa_2\kappa_2$ is frozen according to \eqref{eq:constraint_K}, whereas both $C$ and $\varphi$ are free.

Since 
\begin{equation}\label{eq:s_inequality}
4-\frac{s}{s+1}\frac{1}{\lambda_1^2}>0,
\end{equation}
for all $-1<s_0<1$ and $0<s<1$,
it readily follows that the minimum in \eqref{eq:f_b_iper_reduced} is attained for $\kappa_1^2=\kappa_2^2=|K|$ and for $\varphi$ that minimises 
\begin{equation}
\label{eq:extra_minimum}
\frac{2s}{s+1}\frac{1}{\lambda_1^2}(K-|K|)\sin^22\varphi,
\end{equation} 
that is, assuming $s>0$, for any $\varphi$, if $K>0$ or for $\varphi=\pm\frac\pi4$, if $K<0$. Thus, the minimizers of $f_b$ fall into one of the following two categories:
\begin{enumerate}
	\item\label{case:1} $K\geqq0,\quad \nablas\normal=\pm\sqrt{K}(\n\otimes\n+\nper\otimes\nper)$
	\item\label{case:2} $K\leqq0,\quad\nablas\normal=\pm\sqrt{-K}(\nper\otimes\n+\n\otimes\nper)$
\end{enumerate}
Now we want to see whether there are surfaces $\surface$ and blueprinted fields $\m$ such that these equations are satisfied. Whenever they exist,  they  represent the  energy minimizers in the \emph{vanishing thickness limit}.

In general, a surface $\surface$ (with imprinted $\n$ field) is constructed by solving equations \eqref{eq:connectors_c_d}, as soon as the connectors $\ca$, $\da_1$, and $\da_2$ are known. We know $\ca$ from \eqref{eq:c_star_reduced}, and via \eqref{eq:curvature_tensor_reduced} we can obtain $\da_1$ and $\da_2$ from the knowledge of $\nablas\normal$. It remains to be seen whether these comply with the Codazzi-Mainardi conditions \eqref{eq:skew_part_nabla_d_1}, \eqref{eq:skew_part_d_2}, which are expected to put restrictions on the admissible $\m$ fields.

This study is carried out below for the two categories of minimizers found above. We shall exclude the case where $K\equiv0$, as in this case $\da_1=\da_2=\zero$ and the Codazzi-Mainardi conditions are trivially satisfied for all $\m$, the deformation $\y$ is planar, but not necessarily trivial (if we allow for ridges).
\subsection{Case  $K>0$}\label{sec_case_1}
In this case,
\begin{equation}\label{eq:category_1_d}
\da_1=\pm\lambda_1\sqrt{K}\m,\quad\da_2=\pm\lambda_2\sqrt{K}\mper
\end{equation}
and the Codazzi-Mainardi conditions become
\begin{subequations}\label{eq:Gauss_Mainardi_1}
	\begin{gather}
	\diver\m+\frac{1}{2K}\nabla K\cdot\m=c_2,\label{eq:Gauss_Mainardi_1_1}\\
	\curl\m\cdot\e_3-\frac{1}{2K}\nabla K\cdot\mper=c_1,\label{eq:Gauss_Mainardi_1_2}
	\end{gather}
\end{subequations}
where use has been made of \eqref{eq:c_star_reduced} and of the identity
\begin{equation}\label{eq:identity}
\curl\mper=(\diver\m)\e_3.
\end{equation}
\subsection{Case $K<0$}\label{sec:case_2}
In this case,
\begin{equation}\label{eq:category_2_d}
\da_1=\pm\lambda_2\sqrt{-K}\mper,\quad\da_2=\pm\lambda_1\sqrt{-K}\m
\end{equation}
and the Codazzi-Mainardi conditions take a form very similar to \eqref{eq:Gauss_Mainardi_1},
\begin{subequations}\label{eq:Gauss_Mainardi_2}
	\begin{gather}
	\diver\m+\frac{1}{2K}\nabla K\cdot\m=-c_2,\label{eq:Gauss_Mainardi_2_1}\\
	\curl\m\cdot\e_3-\frac{1}{2K}\nabla K\cdot\mper=-c_1.\label{eq:Gauss_Mainardi_2_2}
	\end{gather}
\end{subequations}
It should be noted that in both \eqref{eq:Gauss_Mainardi_1} and \eqref{eq:Gauss_Mainardi_2} $K$ is meant to be expressed as in \eqref{eq:constraint_K}, so that  these equations are in the end complicated equations for $\m$.

By  \eqref{eq:connector_c_1}, we easily see that 
\begin{equation}\label{eq:identities_c}
\diver\m=c_2,\quad \curl\m=c_1\e_3,
\end{equation}
and so we can give equations \eqref{eq:Gauss_Mainardi_1} and \eqref{eq:Gauss_Mainardi_2} a unified simpler form,
\begin{subequations}\label{eq:Gauss_Mainardi_unified}
	\begin{gather}
	\diver\m+\frac{1}{2K}\nabla K\cdot\m=\pm\diver\m,\label{eq:Gauss_Mainardi_unified_1}\\
	\curl\m\cdot\e_3-\frac{1}{2K}\nabla K\cdot\mper=\pm\curl\m\cdot\e_3,\label{eq:Gauss_Mainardi_unified_2}
	\end{gather}
\end{subequations}
where the upper sign applies to case \ref{case:1} and the lower sign to case \ref{case:2}. Thus, in the former case equations \eqref{eq:Gauss_Mainardi_unified} reduce to
\begin{equation}\label{eq:Gauss_Mainardi_post_unified_1}
\nabla K=0,
\end{equation}
which says that $\surface$ is part of a sphere.\footnote{A result which was to be expected, as in case~\ref{case:1} $\nablas\normal$ is proportional to the projection on the local tangent plane to $\surface$, and so all points of $\surface$ are \emph{umbilics}; the sphere is the only curved surface whose points are all umbilics \cite[p.\,187]{hilbert:geometry}.} In the latter case, with a little more labor, we obtain instead the equation
\begin{equation}\label{eq:Gauss_Mainardi_post_unified_2}
\nabla\ln\sqrt[4]{-K}=\cv_\perp,
\end{equation}
where
\begin{equation}\label{eq:c_perp}
\cv_\perp:=\e_3\times\cv.
\end{equation}
For \eqref{eq:Gauss_Mainardi_post_unified_2} to be integrable, $\nabla\cv_\perp$ must be symmetric, which is guaranteed by the extra condition
\begin{equation}\label{eq:integrability_c}
\diver\cv=0.
\end{equation}
When, as mentioned Sec.~\ref{sec:connectors}, we can write $\cv=\nabla\Phi$, \eqref{eq:integrability_c} amounts to require that the potential $\Phi$ is harmonic.

We have so far failed to find explicit solutions to \eqref{eq:Gauss_Mainardi_post_unified_2} and \eqref{eq:integrability_c}; the argument in Appendix~\ref{sec:remarks} makes us conjecture that there are none. On the other hand, in special cases we have found  fields $\m$ that, once imprinted on the reference configuration, would be lifted on a spherical patch. Below we present the case of a ribbon mapped on a spherical zone.

\subsection{Example}\label{sec:example}
\begin{figure}
	\begin{center}
		\includegraphics[width=.6\linewidth]{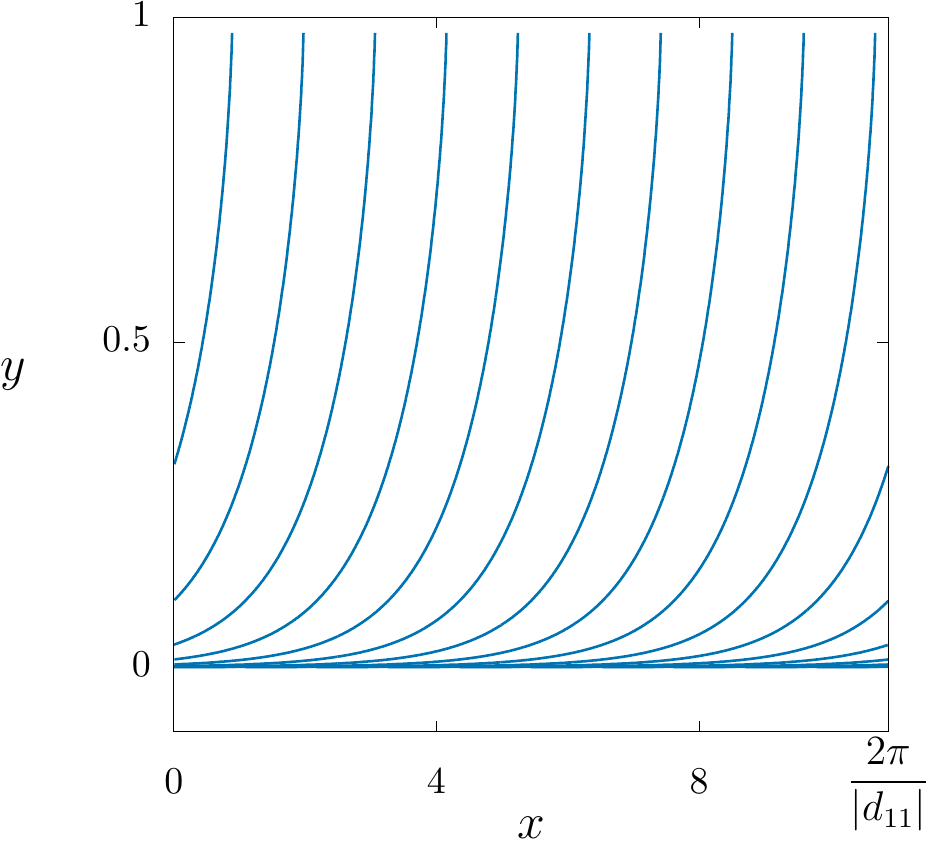}
	\end{center}
	\caption{Integral lines of the field $\m$ in \eqref{eq:m_representation} corresponding to the function $\omega(y)$ in \eqref{eq:solution}. Here $\ell$ is expressed according to \eqref{eq:ell_expression}.}
	\label{fig:integral_lines}
\end{figure}

Consider a ribbon with edges parallel to the coordinate axes of the $(x,y)$ plane. We rescale  lengths to the width along the $y$-axis and we denote by $\ell$ the width  along the $x$-axis, so that  $\ell$ is a dimensionless parameter to be chosen appropriately below. We let $\m$ be represented as 
\begin{equation}
\label{eq:m_representation}
\m=\cos\omega(y)\e_x+\sin\omega(y)\e_y
\end{equation}
and set $\kappa\equiv1$ in \eqref{eq:kappa} (meaning that $K$ is rescaled to the reciprocal of the square of the $y$-edge). It readily follows from \eqref{eq:m_representation} that 
\begin{equation}
\label{eq:c_1_c_2_c_12_representation}
c_1=\omega'\sin\omega,\quad c_2=\omega'\cos\omega,\quad c_{12}=\omega''\sin\omega\cos\omega,
\end{equation}
so that \eqref{eq:kappa} reduces to the differential equation
\begin{equation}
\label{eq:differential_equation}
\frac12\omega''\sin2\omega+\omega'^2\cos2\omega=1,
\end{equation}
which, under the boundary conditions $\omega(0)=0$ and $\omega(1)=\frac\pi2$, is solved by
\begin{equation}
\label{eq:solution}
\omega(y)=\frac12\arccos(1-2y^2).
\end{equation} 
The corresponding integral lines of $\m$ are shown in Fig.~\ref{fig:integral_lines}.
A direct computation shows that for $\omega(y)$ as in \eqref{eq:solution}
\begin{equation}\label{eq:solution_c_1_c_2}
	c_1(y)=\frac{\sin y}{\sqrt{1-y^2}}\quad\text{and}\quad c_2\equiv1.
\end{equation}
Thus, $c_2$ is constant,  $c_1(0)=0$, and $c_1$ diverges to $+\infty$ in $1$.

We shall assume that $s>s_0$, so that $\lambda_1>1$ and $K>0$. Moreover, we shall solve equations \eqref{eq:connectors_c_d} with the initial conditions
\begin{equation}
\label{cond}   
\n(0,0)=\m(0,0)=\e_x,\quad \nper(0,0)=\mper(0,0)=\e_y,\quad \normal(0,0)=\e_z,
\end{equation}
as if we were holding the ribbon clamped by a pair of tweezers in its left bottom corner. 	We now apply both sides of equations \eqref{eq:connectors_c_d} to $\m$ and $\mper$, obtaining
\begin{subequations}
	\label{sysproj}
	\begin{eqnarray}
	(\nabla\n)\m=\lambda_1^2 c_1\nper+d_{11}\normal ,
	&\qquad& 
	(\nabla\n)\mper=\lambda_1^{-2} c_2\nper,
	\\
	(\nabla\nper)\m=-\lambda_1^2 c_1\n,          
	&\qquad& 
	(\nabla\nper)\mper=-\lambda_1^{-2} c_2\n+d_{22}\normal,
	\\
	(\nabla\normal)\m=-d_{11}\n,
	&\qquad& 
	(\nabla\normal)\mper=-d_{22}\nper.
	\end{eqnarray}
\end{subequations}
In particular, since $c_1=0$ and $\m=\e_x$ for $y=0$, from \eqref{sysproj}
we see that along the bottom edge of the ribbon
\begin{subequations}\label{eq:star}
\begin{eqnarray}
\partial_{xx} \n+d_{11}^2\n&=&\bm{0},
\\
\partial_{x}\nper&=&\bm{0},
\\
\partial_x\normal&=&-d_{11}\n,
\end{eqnarray}
\end{subequations}
whose solution, by \eqref{cond}, is
\begin{equation}
\label{y=0}
\n=\cos(d_{11}x)\e_x+\sin(d_{11}x)\e_z,
\quad 
\nper=\e_y,
\quad 
\normal=-\sin(d_{11}x)\e_x+\cos(d_{11}x)\e_z,
\end{equation}
where, in accord with \eqref{eq:category_1_d} and \eqref{eq:constraint_K}, we have chosen
\begin{equation}
\label{eq:d_11_d_22_choices}
d_{11}=-\sqrt{\frac{s-s_0}{1+s_0}}\quad\text{and}\quad d_{22}=-\sqrt{\frac{s-s_0}{1+s}}
\end{equation} 
(changing signs in \eqref{eq:d_11_d_22_choices} would have the effect of reverting the sense of bending, making $\normal$ the \emph{inner} normal to $\surface$). Thus, the bottom edge of the ribbon goes into an arc of circle on the $(x,z)$ plane, which is closed if we choose
\begin{equation}
\label{eq:ell_expression}
\ell=\frac{2\pi}{|d_{11}|}.
\end{equation}
Since $\normal\cdot\e_y=0$, this circle is a great circle of the sphere on which the ribbon is mapped. The radius $R$  of this sphere follows from the equation $\lambda_1\ell=2\pi R$,
\begin{equation}
\label{eq:R}
R=\frac{\lambda_1}{|d_{11}|}.
\end{equation}

We now would like to know whether the system \eqref{sysproj} is periodic, that is, whether with this choice of $\ell$ all other segments at $0<y<1$ close into circles on the sphere of radius $R$.
To this purpose, we compute the $y$ derivative of the frame $(\n,\nper,\normal)$ at fixed $x$ by applying both sides of equations \eqref{eq:connectors_c_d} to $\e_y$,
\begin{subequations}
	\label{sys-y}
	\begin{eqnarray}
	\partial_y\n&=&\omega'(\lambda_1^2\sin^2\omega
	+\lambda_1^{-2}\cos^2\omega)\nper+d_{11}\sin\omega\normal,
	\\
	\partial_y\nper&=&-\omega'(\lambda_1^2\sin^2\omega
	+\lambda_1^{-2}\cos^2\omega)\n+d_{22}\cos(\omega)\normal,
	\\
	\partial_y\normal&=&-(d_{11}\sin\omega\n+d_{22}\cos\omega\nper),
	\end{eqnarray}
\end{subequations}
where use has also been made of \eqref{eq:category_1_d}. This is
is a system of ODEs that propagate the frame $(\n,\nper,\normal)$ at $y=0$ to the interval $0<y<1$, for any given $0\leqq x\leqq\ell$. Two properties can be deduced from \eqref{y=0}. Since the initial values coincide at $x=0$ and $x=\ell$, and \eqref{sys-y} is independent of $x$, all segments parallel to the $x$-edge go into circles. The numerical solutions of \eqref{sysproj} obtained with the algorithm outlined in Appendix~\ref{sec:numerics} are represented in Fig.~\ref{fig-char-sp}.
\begin{figure}[h]
	\centering
	\begin{subfigure}[b]{0.45\textwidth}
		\centering
		\includegraphics[width=\textwidth]{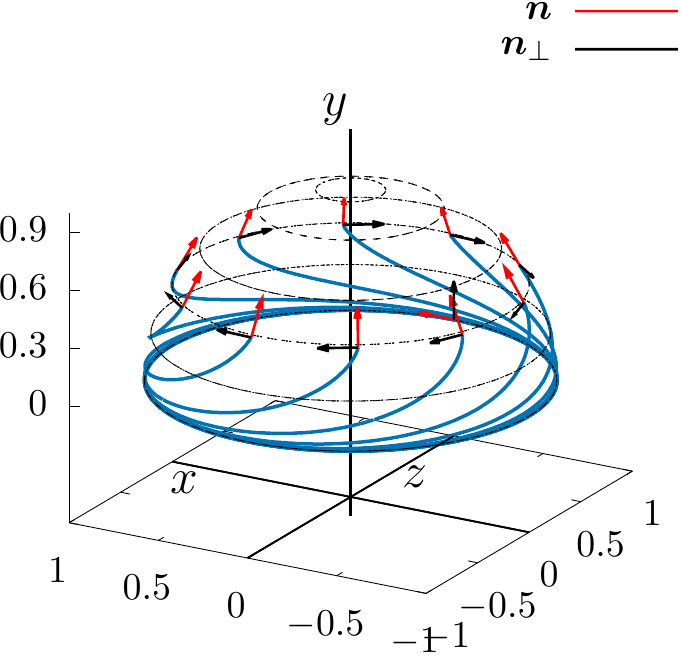}
		\caption{From above}
		\label{fig:view_above}
	\end{subfigure}
	$\quad$
	\begin{subfigure}[b]{0.45\textwidth}
		\centering
		\includegraphics[width=\textwidth]{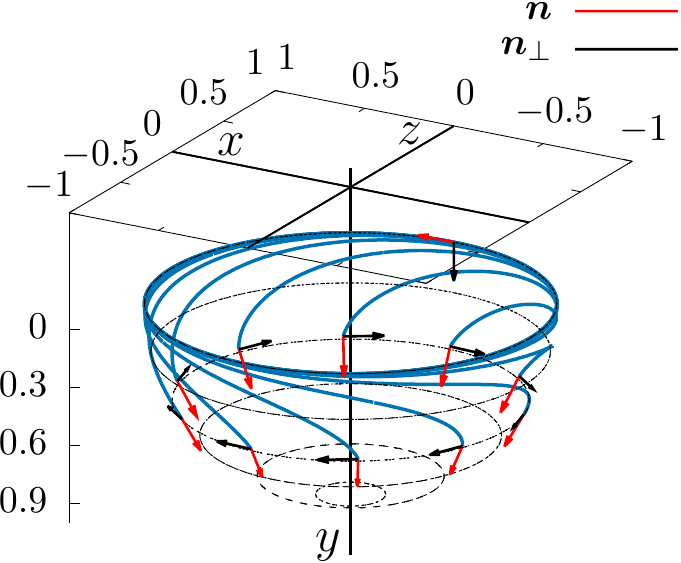}
		\caption{Upside down}
		\label{fig:view_below}
	\end{subfigure}
	\begin{subfigure}[b]{0.4\textwidth}
	\centering
	\includegraphics[width=\textwidth]{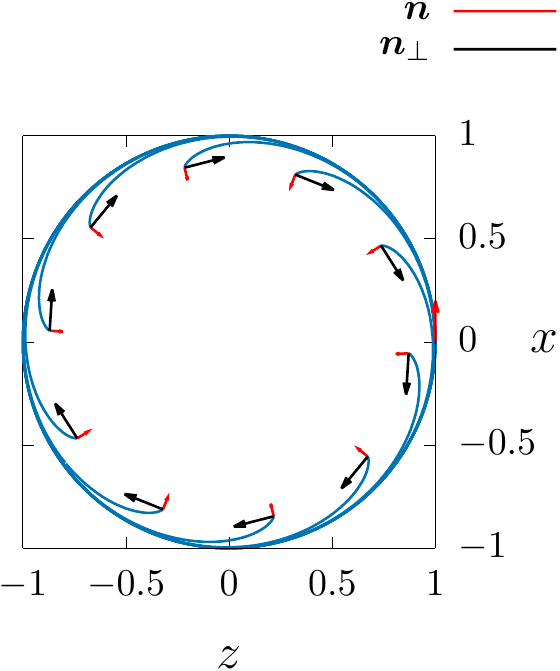}
	\caption{From the top}
	\label{fig:view_top}
\end{subfigure}
$\quad$
\begin{subfigure}[b]{0.5\textwidth}
	\centering
	\includegraphics[width=\textwidth]{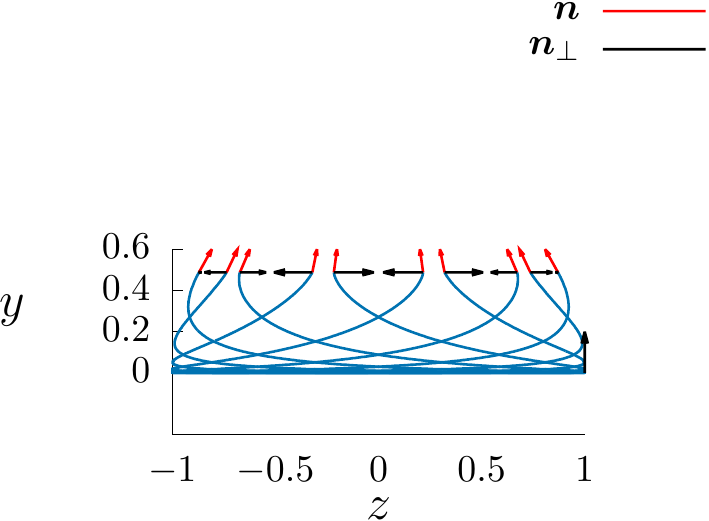}
	\caption{From a side}
	\label{fig:view_side}
\end{subfigure}
	\caption{Four different views in the frame $(\e_x,\e_y,\e_z)$ of the integral lines of $\n$ obtained by deformation of the integral lines of $\m$ shown in Fig.~\ref{fig:integral_lines}. The bottom edge of the ribbon is mapped into a great circle in the $(x,z)$ plane of a sphere of radius $R=\lambda_1/|d_{11}|$, while the top edge is mapped into a parallel of radius $r=1/\lambda_1|d_{11}|$. The deformed ribbon is extended over a meridian amplitude $\theta=\arccos(1/\lambda_1^2)$. In all panels, coordinates are rescaled to $R$ to ensure a better visibility, and $s_0=1/2$, $s=1$; with this choice of parameters, $r/R=1/\lambda_1^2=\sqrt{3}/2.$}
	\label{fig-char-sp}
\end{figure} 

The pictures in Fig.~\ref{fig-char-sp} show that the deformed ribbon does not cover a whole hemisphere, but is confined between a great circle and a parallel, whose radius $r$ we now determine. Making use of the second column of \eqref{sysproj}, since $\mper=-\e_x$ for $y=1$, we get
\begin{subequations}\label{eq:double_star}
\begin{eqnarray}
\partial_{xx} \nper+(\lambda_1^{-4}+d_{22}^2)\nper&=&\bm{0},
\\
\partial_{x}\n&=&-\lambda_1^{-2}\nper,
\\
\partial_x\normal&=&d_{22}\nper.
\end{eqnarray}
\end{subequations}
Since
$\lambda_1^{-4}+d_{22}^2=d_{11}^2$, these equations agree with \eqref{eq:star}. At the same time, since equations \eqref{sys-y} depend only on $y$ and the bottom edge is transformed into a great circle orthogonal to $\e_y$ with unit tangent $\n(x,0)$, $\nper(x,1)$ is generated for all $0\leqq x\leqq\ell$ by rotating $\nper(0,1)$ around $\e_y$; this can be the unit tangent to the deformed top edge only if the latter is a circle parallel to the bottom edge. Thus, integrating \eqref{eq:double_star} we conclude that 
\begin{equation}
\label{eq:upper_edge_equation}
\nper(x,\ell)=-\cos(d_{11}x)\e_x-\sin(d_{11}x)\e_z,
\end{equation}
which is the unit tangent to a circle of radius
\begin{equation}\label{eq:r}
r=\frac{\lambda_2\ell}{2\pi}=\frac{1}{\lambda_1d_{11}},
\end{equation}
so that, by \eqref{eq:R}, $r/R=1/\lambda_1^2$. Correspondingly, elementary trigonometry shows that the deformed ribbon extends along a meridian arc of amplitude $\theta=\arccos(1/\lambda_1^2)$.  

\section{Conclusions}\label{sec:conclusions}
In this paper, we took up the challenge of describing how extrinsic curvatures affect the deformed shape of a thin sheet of nematic polymeric network. We performed a dimension reduction starting from the celebrated neo-classical energy for nematic elastomers in three space dimensions by use of a modified Kirchhoff-Love hypothesis, which was shown to be compatible with more sophisticated analytical methods (relying on $\Gamma$-convergence) \cite{ozenda:kirchhoff}.

The answer was provided by equations \eqref{eq:f_s} and \eqref{eq:f_b}, which deliver the linear and cubic contributions (in the semi-thickness $h$) to the elastic free-energy density (per unit area of the sheet). They embody the stretching and bending energies, respectively. The former, $f_s$, is the one that has been so widely used to predict the (intrinsic) Gaussian curvature $K$ that can be induced by imprinting in the flat reference configuration $S$ of the sheet an appropriate nematic director field $\m$, uniform across its thickness. The latter, $f_b$, is novel and shows how the extrinsic curvatures of the deformed shape $\surface$ feature in the bending: not only through the mean curvature $H$, as expected, but also through an invariant measure depending on the orientation of the conveyed nematic director $\n$ relative to the directions of principal curvature, see $\hat{a}_1$ in \eqref{eq:a_1_hat}. 

The blended energy obtained in Sec.~\ref{sec:blend}, with its stretching and bending components, is different in spirit from a number of rigorous derivations of the bending energy for non-linear plates via $\Gamma$-convergence, as in \cite{friesecke:theorem,friesecke:hierarchy}. There, the bending energy is restricted to isometric immersions, which in our context would correspond to  deformations with prescribed stretching tensor $\C=\C_0$, as in \eqref{eq:C_stationary}. Indeed, in the non-linear theory of plates in \cite{friesecke:theorem,friesecke:hierarchy} only isometric immersions have finite energy, as a result of an intrinsic limitation of the analytic technique employed there, which (at least until this time) is unable to provide limiting energies depending on the semi-thickness $h$. Said differently, apart from some attempts\footnote{Not even very satisfactory, as appropriately remarked in \cite{friesecke:hierarchy}: ``It seems that these $\Gamma$-expansions tend to separate the regimes more than is desirable; each successive term in the $\Gamma$--expansion can only make an arbitrarily small perturbation to the preceding theory.''} in linear settings \cite{anzelotti:dimension}, $\Gamma$-convergence has not yet evolved into $\Gamma$-expansion.

In contrast, our blended energy \eqref{eq:energy_blended} depends on $h$ (up to the third order) and applies to all deformations; as better shown by \eqref{eq:f_b}, the stretching tensor $\C$ also features in the bending energy $f_b$. In the presence of different boundary conditions or different imprinted nematic fields, one could expect either the stretching or the bending energy to characterize the response of the material in different regions of a thin sheet; equilibria will result from a balance between the two blended components of energy.

The elastomeric shells considered in this paper were homogeneous across the thickness; the director field imprinted in the reference configuration was uniform in the $x_3$ coordinate. Inspired by an extension to laminates \cite{schmidt:plate} of the non-linear plate theory in \cite{friesecke:theorem,friesecke:hierarchy}, a number of recent works   have derived a bending energy for nematic elastomer shells with an imprinted  director varying across the thickness \cite{agostiniani:rigorous,agostiniani:shape,agostiniani:dimension}. Remarkably, the obtained energies are identical in structure with those posited by geometric elasticity \cite{klein:shaping,aharoni:geometry,armon:geometry}. As in the parent theory \cite{schmidt:plate}, these energies are appropriate $\Gamma$-limits as $h\to0$ and are valid on isometric immersions. It would be desirable to obtain also for these non-homogeneous elastomer sheets a blended energy like \eqref{eq:energy_blended}, where $h$ features explicitly.

We also studied the vanishing thickness limit of the blended energy, in which, with a two-step strategy, $f_b$ is minimized over the minimizers of $f_s$, thus providing a selection criterion for the multitude of shapes that would result from minimizing only the stretching energy. We showed that this strategy is successful when the (prescribed) Gaussian curvature $K$ is positive (actually, constant), in which case we gave an explicit construction. We also conjectured (with some supporting argument) that this strategy fails if $K<0$.

Of course, the real challenge is now to minimize the whole surface elastic energy, in which $f_s$ and $f_b$ are blended together, and  not necessarily scale-separable. Appropriate boundary conditions should be imposed on $\partial S$ in the case of spontaneous deformations (actuated, for example, by heat or light), which may prescribe either position or outer unit tangent. We would expect that these problems could only be solved numerically, for which task we need to develop appropriate algorithms. We do hope that having explicit (and not too complicated) formulae for $f_s$ and $f_b$ as in \eqref{eq:f_s} and \eqref{eq:f_b} would encourage researchers to take up the non-trivial challenges offered by these problems.

\begin{acknowledgments}
Most of the contents of this paper were fist illustrated by E.G.V. in a lecture given in December 2019  at the Institute for Computational and Experimental Research in Mathematics (ICERM) in Providence, RI, during the Workshop on \emph{Numerical Methods and New Perspectives for Extended Liquid Crystalline Systems}, see  \url{https://icerm.brown.edu/video_archive/?play=2112}. The kindness of the organizers of the Workshop and the generous hospitality of ICERM are gratefully acknowledged. 
The work of O.O. was supported financially by the Department of Mathematics of the University of Pavia as part of the activities funded by the Italian MIUR under the nationwide Program ``Dipartimenti di Eccellenza (2018-2022).''
\end{acknowledgments}

\appendix
\section{Derivation of equation \eqref{eq:bulk_energy_density} from the trace formula}\label{sec:derivation}
Here we shall sketch the steps that led us from \eqref{eq:energy_density} to \eqref{eq:bulk_energy_density}. First, from \eqref{eq:L_n} we compute
\begin{equation}
\label{eq:L_n_inverse}
\Ln^{-1}=\frac{1}{a}\left(\I -\frac{s}{1+s}\m\otimes\n\right).
\end{equation}
Making use of \eqref{eq:L_n_inverse} and \eqref{eq:L_m}, we obtain that 
\begin{equation}
\label{eq:derivation_b}
\Ln^{-1}\F\Lm=\frac{a_0}{a}\left(\F+\frac{s_0}{s+1}\F\m\otimes\m-\frac{s}{s+1}\frac{1}{\m\cdot\Cf\m}\F\m\otimes\Cf\m \right),
\end{equation}
where resort has been made to \eqref{eq:n} and to the identity
\begin{equation}
\label{eq:identity_c}
|\F\m|^2=\m\cdot\Cf\m.
\end{equation}
It is now a simple matter to arrive from \eqref{eq:derivation_b} to 
\begin{equation}\label{eq:derivation_d}
\F\trans\Ln^{-1}\F\Lm=\frac{a_0}{a}\left(\Cf+\frac{s_0}{s+1}\Cf\m\otimes\m-\frac{s}{s+1}\frac{1}{\m\cdot\Cf\m}\Cf\m\otimes\Cf\m \right).	
\end{equation}
Taking the trace of both sides of \eqref{eq:derivation_d}, we easily retrace \eqref{eq:bulk_energy_density} in the text, which, unlike \eqref{eq:energy_density}, no longer features $\n$ explicitly, as a consequence of the freezing constraint \eqref{eq:n}.

\section{No solution with negative Gaussian curvature?}\label{sec:remarks}
Here we sketch an argument that makes us conjecture that equations \eqref{eq:Gauss_Mainardi_post_unified_2} and \eqref{eq:integrability_c} have no solution.
Letting $\m=\cos\vt\e_x+\sin\vt\e_y$, with $\vt=\vt(x,y)$, we write them as
\begin{subequations}\label{eq:kappa_equations}
	\begin{gather}
	\nabla\kappa=4\kappa(\vt_{,x}\e_y-\vt_{,y}\e_x),\label{eq:nabla_kappa}\\
	\nabla^2\vt=0,\label{eq:laplacian_theta}
	\end{gather}
	where, by \eqref{eq:kappa},
	\begin{equation}\label{eq:kappa_theta}
	\kappa=\cos2\vt(\vt_{,xy}+\vt_{,y}^2-\vt_{,x}^2)+\sin2\vt(\vt_{,yy}-2\vt_{,x}\vt_{,y})\neq0.
	\end{equation}
\end{subequations}
It follows from \eqref{eq:nabla_kappa} that 
\begin{equation}\label{eq:consequence_1}
\nabla\kappa\cdot\nabla\vt=0,
\end{equation}
and from \eqref{eq:laplacian_theta} that 
\begin{equation}\label{eq:consequence_2}
\int_{\curve}\vt\nabla\vt\cdot\normal_\curve\dd s=\int_{\region}|\nabla\vt|^2\dd a,
\end{equation}
for any closed, regular curve $\curve$ in the $(x,y)$ plane enclosing the region $\region$
with outer unit normal $\normal_{\curve}$. 

Now, we show that no curve $\curve$ can be a level set of $\kappa$.  If $\kappa$ were  constant on $\curve$, then $\nabla\kappa=\lambda\normal_\curve$ and, if  $\lambda\neq0$, \eqref{eq:consequence_1} would require $\nabla\vt\cdot\normal_S$ to vanish on $\curve$, otherwise \eqref{eq:nabla_kappa} would require $\nabla\vt$ to vanish on $\curve$. In both instances, \eqref{eq:consequence_2} implies that $\nabla\vartheta$ vanishes in $\region$, making  $\kappa\equiv0$, by \eqref{eq:kappa_theta}, a contradiction. Thus, non trivial solutions of \eqref{eq:kappa_equations} must have all level sets for $\kappa$ that hit the boundary of the reference domain $S$. We somehow feel that this is too restrictive a condition to be met, but we lack a proof to substantiate our conjecture that non-trivial solutions of \eqref{eq:kappa_equations} do not exist.

\section{Numerical algorithm for lifting a decorated ribbon}\label{sec:numerics}
	In this appendix, we show how to solve numerically the first column of equations \eqref{sysproj}. We formulate them as Gateau derivatives,
	\begin{subequations}
		\label{sys-char}
		\begin{eqnarray}
		\lim_{h \to 0} \frac{\n((x,y)+h\m(x,y))-\n(x,y)}{h}
		&=&
		\lambda_1^2 c_1\nper+d_{11}\normal,
		\\
		\lim_{h \to 0} \frac{\nper((x,y)+h\m(x,y))-\nper(x,y)}{h}
		&=&
		-\lambda_1^2 c_1\n,          
		\\
		\lim_{h \to 0} \frac{\normal((x,y)+h\m(x,y))-\normal(x,y)}{h}
		&=&
		-d_{11}\n.
		\end{eqnarray}
	\end{subequations}
	For $(x,y)\in(\mathbb{R}/\ell\mathbb{R}\times[0;1[)$ and $t>0$, let $X((x,y),t)$ be the integral curves associated with the vector field $\m$. They satisfy  $X((x,y),0)=(x,y)$, and 
	for $t>0$, 
	\begin{equation}
	\label{eq-flat-char}
	\frac{\partial X(x,t)}{\partial t}=\m(X(x,t)).
	\end{equation} 
	Equation \eqref{eq-flat-char} is solved with an explicit Euler scheme and a time step $\Delta_t=10^{-4}$
	for initial value $(x,y)=(0,\Delta_t/2)$. The final time $t_f$ is reached when $X((x,y),t_f)\cdot\e_y=0.98$. 
	As $\m$ does not depend on $x$, integral lines given by the set of 
	values $\ x_i=\left(\frac{i\ell}{10}\right)_{0\le i\le 10}$ 
	are obtained by horizontal translation. 
	The result is shown in Fig.~\ref{fig:integral_lines}. 

	Let $X_{i,k}$ denote  the approximate value of $X((x_i,\Delta_t/2),k\Delta_t)$. 
	The following explicit numerical method is applied:
	\begin{eqnarray*}
		\n(X_{i,k+1})-\n(X_{i,k})
		&=&
		\Delta_t(\lambda_1^2 c_1(X_{i,k})\nper(X_{i,k})+d_{11}(X_{i,k})\normal(X_{i,k})),
		\\
		\nper(X_{i,k+1})-\nper(X_{i,k})
		&=&
		-\Delta_t \lambda_1^2 c_1(X_{i,k})\n(X_{i,k}),          
		\\
		\normal(X_{i,k+1})-\normal(X_{i,k})
		&=&
		-\Delta_t d_{11}(X_{i,k})\n(X_{i,k}).
	\end{eqnarray*}
	The result is shown in Fig.~\ref{fig-char-sp}, where theory was used for guidance in rounding off the integral lines of $\n$ in the vicinity of the bottom and top circles.


%

\end{document}